\documentclass[12.pt]{article}
\input{epsf}

\begin{document}

\bibliographystyle{unsrt}

\centerline{ \huge \bf Circles, Spheres and Drops Packings}

\vskip.7cm \centerline{Tomaso Aste}

\centerline{Laboratoire de Physique Th\'eorique, Universit\'e Louis Pasteur, 3 rue de
l'Universit\'e, } \centerline{67084 Strasbourg, France and C.I.I.M., Universit\`a di Genova,
Piazz.  Kennedy Pad.  D,16129 Genova, Italy.}

\vskip2.cm {\large \bf Abstract} \vskip.7cm \noindent We studied the geometrical and topological
rules underlying the dispositions and the size distribution of non--overlapping, polydisperse
circle--packings.  We found that the size distribution of circles that densely cover a plane
follows the power law:  $N(R) \propto R^{-\alpha}$.  We obtained an approximate expression which
relates the exponent $\alpha$ to the average coordination number and to the packing strategy.
In the case of disordered packings (where the circles have random sizes and positions) we found
the upper bound $\alpha_{Max} = 2$.  The results obtained for circles--packing was extended to
packing of spheres and hyper--spheres in spaces of arbitrary dimension $D$.  We found that the
size distribution of dense packed polydisperse $D$--spheres, follows --as in the two dimensional
case-- a power law, where the exponent $\alpha$ depends on the packing strategy.  In particular,
in the case of disordered packing, we obtained the upper bound $\alpha_{Max}=D$.
Circle--covering generated by computer simulations, gives size distributions that are in
agreement with these analytical predictions.  Tin drops generated by vapour deposition on a hot
substrate form breath figures where the drop--size distributions are power laws with exponent
$\alpha \simeq 2$.  We pointed out the similarity between these structures and the
circle-packings.  Despite the complicated mechanism of formation of these structures, we showed
that it is possible to describe the drops arrangements, the size distribution and the evolution
at constant coverage, in term of maximum packing of circles regulated by coalescence.

\vskip2.cm \section{Introduction} The plane is not very efficiently covered with circles, but
circle--covering is a good model--system for the study of the formation and evolution of many
natural and artificial systems as granular materials \cite{Bide91,Guy90}, island formation in
metal films \cite{Black91,Black94}, segregation problems \cite{Troad94}, plate tectonics and
turbulence \cite{Herr90,Herr91}, focal arrangement in smetic liquid crystals \cite{Bou72,Bid73}.
One interesting problem related with packing circles is to find the size distribution that
efficiently leads to the densest covering, compatible with the packing strategy.  In some
particular cases, as the Apollonian covering \cite{Cox61}(where the circles are packed all
tangent to each other following a defined sequence), analytical and numerical solution are
available in literature \cite{Herr90,Herr91,Boyd73,Mand82}.  In this paper the problem is
discuss in the general case of disordered packing (or ``osculatory'' packing), where the circles
are set with random sizes and positions.  The motivation of this work is understand the basic
mechanisms which are underlying the morphogenesys of breath figures
\cite{Bey86,Bri90,Fri90,Kno88,Fam88,Bey90I}.  Water which condense on a glass form a
densely-packed system of droplets known as breath figures.  The formation of the droplet system
is regulated by two basic mechanisms:  the {\it independent grow} of a drop supported by the
condensation and the melting of two or more drops in a {\it coalescence} phenomena.  These two
mechanisms leads to systems with a wide range in the drop sizes.  The distribution is
characterized by rather uniform drop sizes in the region of large radii (drops which are grown
from the originally firstly nucleated drops trough a chain of independent grow and coalescence)
and a power law in the region of small radii (drops which are generated by re-nucleation on the
surface liberated by coalescence).  The a formation of these structures is an extremely
complicate dynamical process.  The point of view here adopted is that the system-morphogenesys
is mainly driven by the geometrical constrains which regulate the drop-packing.

Two dimensional structures generated by packing circles (or by packing others isotropic natural
objects like dew drops on a glass), have some properties which are independent of the specific
formation mechanism.  The main similarities can be schematized in the following three points:
(i) wide range in the sizes; (ii) scale invariance in the packing arrangement; (iii) power law
in the size--distribution ($N(R) \propto R^{-\alpha}$).  Let us briefly discuss about the
possible physical origin of these three similarities:

\begin{itemize}

\item[i)] A wide range in the particle sizes is a necessary condition to reach dense packing.
For example, the packing of circles with equal radii $R_0$ has a maximum density of $0.907$.
The density can be raised to $\rho=0.95$ by filling the interstitial spaces with circles of
sizes $R_1=R_0/6.4$.  Another increment to $\rho=0.97$, requires interstitial radii equal to
$R_2=R_0/15.9$.  In general, the density can be increased up to any value $\rho<1$, but this
requires the utilization of interstitial circles with dimensions which rapidly decrease.

\item[ii)] Consider a procedure where a dense circle-packing is generated by filling the
interstitial spaces with circles of maximum sizes compatibly with the condition of non
overlapping.  (Note that, fixed the maximum radius, this procedure generates structures which
minimizes the surface extension at fixed covered area.)  In this procedure the only relevant
metric parameter is the ratio between the external radii of the circles that generate an
interstice and the internal radius of the circle that fill this interstice.  For example, in the
Apollonian case, this ratio tends to the value $x \simeq 2.9$.  In general, $x$ tends to a
constant value that is related to the packing strategy.  It is straightforward to see that
packing characterized by constant values of $x$ are scale--invariant.

\item[iii)] The number of circles introduced at any covering step increases following a
geometrical progression.  For example, in the Apollonian packing, if one starts with four
circles in contact, at the first covering step one has 3 interstitial spaces to fill by circles,
at the second step the interstices to fill are 9, at the third step this number is 27 and, at
the $\nu^{th}$ are $3^{\nu}$.  In the general case, the number of circles introduced at any
stage grows as a geometrical progression $a^\nu$.  By associating this increment in the number
of interstitial circles with the condition of scale invariance, one obtains a size distribution
that follows the law $N(R) \propto R^{-\alpha}$, where the coefficient $\alpha$ depends on the
packing strategy.  \end{itemize}

In this paper we obtain an approximate expression that relates the exponent $\alpha$ to the
local topological properties of the circle-packing and to the packing strategy.  The limiting
value of such coefficient in the case of disordered maximum packing have been found equal to 2.

Simple computer simulations of disordered circle covering have been performed.  The results,
presented in section \ref{S.4}, confirm the theoretical predictions:  the circle size
distribution follows a power law with exponent $\alpha \simeq 2$.

The size distribution of tin droplets condensed on a hot, flat surface have been experimentally
studied \cite{AstePhD,PozTes,SayTes,AsBoBud,Aste/Bot94}.  In section \ref{S.5}, the experimental
data are interpreted in terms of maximum circles packing.  Through this point of view it is
possible to give a simple explanation of some statistical and dynamical characteristic of these
systems.  In particular, the drop size distribution, the coverage and the self similar evolution
at constant coverage, have been interpreted in the framework of maximum packing of drops
regulated by coalescence.

\section{\label{S.2} Topological and geometrical rules in packing }

\begin{figure}
\vspace{-1.cm}
\hspace{4.cm}
\epsfxsize=18.cm
\epsffile{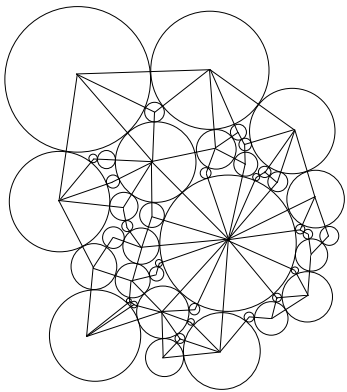}
\label{f.1}
\vspace{-5.cm}
\caption{The Dodds network generated by conneting the centers of circles mutually in contact. 
}
\end{figure}

Consider a plane densely covered by circles placed at random following the only constraint of
non overlapping.  Consider the Dodds network \cite{Dodds80} constituted by the edges that
connect the centres of circles in contact (see.  fig.(\ref{f.1}) ).  Such a network has a number
of vertices (in the following indicated by $V$) equal to the number of circles.  The vertex
connectivity ($\langle z \rangle$) is the average number of neighbours in contact with a circle.
The other network parameters (in particular the number of edges $E$ and the number of faces $F$)
are given, in term of $V$ and $\langle z \rangle$, by the following relation \cite{Bid86}
\begin{equation} \langle z \rangle V=2E \label{connect} \end{equation} (any vertex is surrounded
by $\langle z \rangle$ edges and any edge is bounded by 2 vertices\footnote{This relation holds
only if in the network there aren't insulated vertices.  This condition is generally satisfied
in dense packed systems.}), and by the Euler's formula 
\begin{equation} V-E+F=\chi \;\;\;\; ,
\label{Eul} 
\end{equation} 
where $\chi$ is the Euler Poincarr\'e characteristic and takes the
value of 1 for an Euclidean plane.

The combination of equations (\ref{connect}) and (\ref{Eul}) leads to a useful relation between
the number of circles $V$ and the number of interstitial spaces between the circles (that is
equal to the number of faces ($F$) of the Dodds network), we have:  \begin{equation} F=\left(
{\displaystyle \langle z \rangle \over 2}-1 \right)V+\chi \;\;\; .  \label{n_inter}
\end{equation}

Note that at the point $\langle z \rangle=2$ the interstitial spaces between circles become
connected ($F=\chi = 1$, in the Euclidean plane) and, consequently, the network unconnected.
This is the percolation threshold in two dimension.  For the purposes of the present paper it is
interesting the opposite limit (as far as we are investigating the properties of dense packing).
From a topological point of view, the densest circle packing is characterized by circles which
are all in contact between neighbours.  In such a packing the interstices between circles are
all closed and surrounded by three circles.  In this case, holds the identity $3F=2E$ (any face
has three edges and any edge divides two faces) and the vertex connectivity reach its maximum
value:  $\langle z \rangle_{Max}=6 (1-{\chi \over V} )$.  In general, less dense packing are
characterized by lower value in the vertex connectivity (for example a typical connectivity for
disordered packing of binary and polydisperse mixtures of circles is $\langle z \rangle \simeq
3.75$ \cite{Bid86}).

\subsection{\label{S.2.1} Apollonian covering}

An example of topologically densest circle packing is the Apollonian covering.  In such a
packing, any circle is in contact with 3 surrounding circles.  A well known formula ({\it The
kiss precise} \cite{Cox61}, \cite{Sod36}) gives a relation between the bends $ \epsilon_i=1/R_i$
of the four ``kissing'' circles of radii $R_i$, $i=1,...4$ \begin{equation}
2(\epsilon_1^2+\epsilon_2^2+\epsilon_3^2+\epsilon_4^2)
=(\epsilon_1+\epsilon_2+\epsilon_3+\epsilon_4)^2 \;\;\;.  \label{bends} \end{equation}

Following Coxeter \cite{Cox68}, one can consider the case in which the bends belongs to a
geometric progression $ \epsilon_{\nu} = x^{\nu} \epsilon_0$, where $x$ is the ratio of the
progression.  Substituting into (\ref{bends}) one gets a sixth order equation in $x$ with only
one real root bigger than one:  $x \simeq 2.89$.  In general, similar values for the ratio
$\epsilon_{\nu+1}/\epsilon_\nu$ are characteristic of several covering sequences.  For example,
a value slightly less than 2.9, is the ratio at which converge the loxodromic spiral sequence
\cite{Cox68}.  These values corresponds to particular sequences selected from the whole covering
set.  The analytical evaluation of an average value of $x$ corespondent to the complete set of
covering sequences is still an open problem \cite{RivPC}.

The Apollonian covering is a well-defined problem and constitute an excellent example of
circle-covering.  On the other hand, the aim of this work is the study of natural systems which
have an higher degree of randomness in the packing strategy.

\subsection{\label{S.2.2} Disordered covering}

Let us consider a covering procedure that utilizes non overlapping circles with sizes
distributed between two radii $R_{Max}$ and $R_{min}$.  Such a procedure starts filling the
available space with the circles of higher radius $R_0=R_{Max}$.  Once the interstitial space
between these circles become too small to contain circles with maximum radius the procedure
reduces the radius of the new circles to insert $R_1=R_{0}/x$.  The filling continues iterating
this process gradually reducing the radii $R_\nu=x^{-\nu}R_0$ in order to insert circles of
biggest radii compatibly with the interstitial sizes.  Placing the centres in appropriate
positions, this procedure generates the Apollonian sequences of tangent circles discussed in the
previous paragraph.  More generally, one can study a disordered covering where the circles are
placed with the centres at random.  Differently from the Apollonian case, in this covering the
circles are not all tangent.  In the Apollonian covering the interstitial space is always
generated by three circles all in contact with each other.  In the disordered case, the same
three circles are --in general-- not all tangent and the interstitial space is bigger.  It
follows that, in this case, the interstitial space can be filled with circles of bigger sizes
than in the Apollonian case.  Consequently, the ratio between the radii of the circles that form
the boundary of an interstice and the circle that fills this interstice is lower in the
disordered covering respect to the Apollonian case.  Thus, the value $x \simeq 2.9$ suggested as
a limit of the ratio $\epsilon_{\nu+1}/\epsilon_\nu$ in the Apollonian sequences of tangents
circles, represents, for the disordered case, an upper limit.

It is now interesting to investigate the lower limit for $x$.  This limit can be achieved by
imposing the condition that the sum of the areas of all the circles introduced with the covering
procedure converges to a finite value (which must be lower than the total area to cover).

Consider, as before, a covering characterized by a sequence of radii in the geometrical
progression $\epsilon_\nu^{-1}=R_\nu=x^{-\nu}R_0$.  The number of interstitial circles
introduced at any stage $\nu$ is equal to the number of interstices of the system at that stage.
This number is given by eq.(\ref{n_inter})

\begin{equation} F_\nu=\left( {\displaystyle \langle z \rangle \over 2}-1 \right)V_\nu+\chi
\;\;\; , \label{n_inter1} \end{equation} where $V_\nu$ is the number of circles at the stage
$\nu$.

The covering goes on, filling these interstices with $F_\nu$ new circles of radius $R_{\nu+1}$.
The number of interstices at the stage $\nu+1$ is then \begin{equation} F_{\nu+1}=\left(
{\displaystyle \langle z \rangle \over 2}-1 \right) \left(V_\nu+F_\nu \right)+\chi \;\;\; .
\label{n_inter2} \end{equation}

Substituting the value of $V_\nu$ given by eq.(\ref{n_inter1}) into (\ref{n_inter2}), one gets
\begin{equation} F_{\nu+1}=\left( {\displaystyle \langle z \rangle \over 2}\right) F_\nu \;\;\;
, \label{F_seq} \end{equation} that yields \begin{equation} F_{\nu}=\left( {\displaystyle
\langle z \rangle \over 2}\right)^\nu F_0 \;\;\; .  \label{F_seq1} \end{equation}

The total area covered by the circles at the stage $\nu$ is given by the sum over the areas
covered at any stage until $\nu$ \begin{equation} A_\nu=\pi \left(V_0R_0^2+ \sum_{i=1}^\nu F_i
R_i^2 \right) = \pi R_0^2 \left( V_0+F_0{ \langle z \rangle \over 2x^{2}} \sum_{i=0}^\nu \left(
{\displaystyle \langle z \rangle \over 2 x^2}\right)^i \right) \;\;\; .  \label{Area}
\end{equation}

When $\nu \rightarrow \infty$, this sum converge to a finite number only if ${\langle z \rangle
/ (2 x^2)}<1$.  This condition gives the lower limit for the sequence ratio \begin{equation}
x>\sqrt{\langle z \rangle \over 2}\;\;\;.  \label{xx} \end{equation}

Note that eq.(\ref{F_seq1}) have been obtained supposing the coordination number $\langle z
\rangle$ and the ratio $x$ being constant at any covering stage.  This is not in general true.
(The value of $\langle z \rangle$ is fixed and equal to 6 in the Apollonian covering of an
infinite Euclidean plane but not in general.  Moreover --also in this particular case-- the
ratio $x$ is constant only in the asymptotic limit.)  Should be noted that, the convergence
condition for the total area, constrains only the asymptotic values of $x$ and $\langle z
\rangle$.  (It is straightforward to see that, in the Apollonian case, equation (\ref{xx}) is
satisfied by the asymptotic values ($x \simeq 2.9$ and $\langle z \rangle=6$).)  In the
asymptotic limit, the covering procedure is in general scale-invariant, as the only important
metric parameter is the ratio between the sizes of the circles that generate an interstice and
the size of the circle that fills this interstice.  In this asymptotic limit the parameters
$\langle z \rangle$ and $x$ are related only to the covering strategy and are independent of the
specific stage $\nu$.  The inequality (\ref{xx}) constrains these asymptotic parameters.

\section{ \label{S.3} Circle sizes distribution }

The circle--size distribution ($N(R)$) is related to the value of $x$, and to the number of
cells introduced at any covering step.  By associating the relation $R_\nu=R_0 x^{-\nu}$ with
eq.(\ref{F_seq1}), one gets \begin{equation} N(R_\nu)=N_0 \left( {\displaystyle R_0 \over
R_\nu}\right)^\alpha \;\;\; , \label{distrib} \end{equation} where the exponent is
\begin{equation} \alpha={\displaystyle \ln \left( {\displaystyle \langle z \rangle \over 2}
\right) \over \ln x} \;\;\; .  \label{exponent} \end{equation}

Above we pointed out that the value of $\langle z \rangle$ is associated with the packing
strategy:  it is equal to 6 in the Apollonian case, whereas in the disordered covering it tends
gradually to asymptotic values lower or equal than 6.  In the previous paragraph we discussed
the bounds on the value of $x$.  Substituting the lower bound given by eq.(\ref{xx}) one gets
the upper limit for the exponent $\alpha$, \begin{equation} \alpha < {\displaystyle \ln \left(
{\displaystyle \langle z \rangle \over 2} \right) \over \ln \sqrt{\displaystyle \langle z
\rangle \over 2}} = 2 = \alpha_{Max} \;\;\; , \label{alp_lim} \end{equation} which is
independent on $\langle z \rangle$.

We discussed above that the upper bound of $x$ is associated with the Apollonian covering.  In
this case we have $x \simeq 2.9$ and $\langle z \rangle=6$ which leads to $\alpha=1.03$.

The Haudorff Basicovitch fractal dimension of the packing can be directly derived from the size
distribution \cite{Herr90}, one finds $d_f=\alpha$.  The analytical determination of $d_f$ is a
challenging problem of surprising difficulty.  Only bounds are known $1.300197 < d_f < 1.314534$
\cite{Herr90,Boyd73}.  The value here derived is quite far from the two exact bounds and from
the value $d_f=1.305684$ obtained by numerical calculations.  It should be noted that our
relation (\ref{exponent}) is an approximate solution and that the value of $x=2.9$ that we
assumed in (\ref{alp_lim}) is only an estimation valid for a particular subset of all the
possible covering sequences.  On the other hand, values of $1 \le \alpha \le 2$ are in agreement
with two exactly solvable models discussed in appendix \ref{A.2} \cite{Bid73}.  In these
particular cases (where $x$ is constant) one have $d_f=\alpha=1$ and $d_f=\alpha=1.585$.

Note that, in literature (for example \cite{Herr90,Boyd73}) the size distributions are discussed
in the continuos limit.  Here $N(R_\nu)$ is a discrete distribution stating the number of
circles with radius $R_\nu$.  In the continuum limit, $N(R_\nu)$ corresponds to the integral of
the continuous distribution $n(R)$ evaluated between $R_{\nu+1}$ and $R_{\nu}$.  From
eq.(\ref{distrib}) and using the identity $R_{\nu+1}=R_{\nu}/x$, one obtains $n(R)=C
R^{-\alpha-1}$ with $C=\alpha N_0 R_0^\alpha/(x^\alpha -1)$ ($n(R)dR$ being the number of
circles with radius between $R$ and $R+dR$).

\subsection{\label{S.3.1} Generalization to arbitrary dimensions}

It is possible to extend the results obtained in the previous paragraph spaces of any dimension.
The calculus can be done following the same principal points utilized in the two dimensional
case, extending the notions developed for the circles to $D$--spheres.

Consider the network obtained by connecting with edges the centres of all the $D$--spheres
mutually in contact.  The number of vertices ($V$) of such a network is equal to the number of
$D$--spheres.  The number of cells (in the following indicated with $C$) is given by the
relation \begin{equation} C={\displaystyle n_{D,0} \over n_{0,D} } V \;\; , \label{inc}
\end{equation} where $n_{D,0}$ is the mean number of $D$--cells incident on a vertex and
$n_{0,D}$ is the mean number of vertices bounding a $D$--cell.  In two dimension, the incidence
number $n_{2,0}$ is the vertex connectivity $\langle z \rangle$ and $n_{0,2}$ is the mean number
of edges per face $<n>$.  In a planar network these two parameters are connected trough the
relation:  \begin{equation} \langle n \rangle={\displaystyle 2\langle z \rangle \over (\langle z
\rangle-2+{2 \chi \over V})} \;\;.  \label{n/z} \end{equation} Substituting this relation into
(\ref{inc}) one gets immediately eq.(\ref{n_inter1}).  For $D>2$, $n_{D,0}$ and $n_{0,D}$ are,
in general, independent, free parameters related to the network peculiarities.\footnote{For
example, a particular case is the topological densest packing where any $D$--sphere is in
contact with all the surrounding spheres.  In this case any interstitial space is bounded by
$D+1$ $\;D$--spheres.  In such a packing the network is made only by hyper--tetrahedron and one
has $n_{0,D}=D+1$ (any hyper--tetrahedron has $D+1$ vertices).  On the other hand, the incidence
number $n_{D,0}$ is --also in this particular case-- a free parameter which depends on the
network characteristic.  For $D=3$, the maximum value of $n_{3,0}$ for the packing of equal
spheres is 22.6 \cite{Mos/Sad91}.  By filling the interstices with lower sized spheres this
number is reduced up to the lower limit of 12.}

Consider the system at the covering stage $\nu$.  At this stage, $V_\nu$ $D$--spheres are
packed.  The number of interstices between them is $C_\nu=(n_{D,0}/n_{0,D}) V_\nu $.  At the
following stage $\nu+1$, these interstices are filled with $C_\nu$ $\;D$--spheres.  Thus the
total number of $D$--spheres in the system become \begin{equation} V_{\nu+1}=V_\nu+C_\nu=\left(
1+{\displaystyle n_{D,0} \over n_{0,D} } \right) V_\nu\;\;\;.  \label{totn} \end{equation}
Consequently the number of interstices is \begin{equation} C_{\nu+1}=\left(1+{\displaystyle
n_{D,0} \over n_{0,D} }\right)C_\nu \;\; , \label{cc} \end{equation} that gives \begin{equation}
C_\nu=\left(1+{\displaystyle n_{D,0}\over n_{0,D}}\right)^\nu C_0 \;\;\; .  \label{recur}
\end{equation} This equation generalizes eq.(\ref{F_seq1}).

The covering procedure should be scaling invariant:  in the covering strategy it is only
important the ratio between the radii of the $D$--spheres that bound the interstitial space and
the radius of the sphere that fills this interstice.  The only important metric parameter that
characterizes the covering procedure is the ratio $x=R_\nu/R_{\nu+1}$.  A limiting value for $x$
can be obtained --as in the 2$D$ case-- imposing that the sum over the volumes, introduced with
the covering procedure, converges to a finite value:  \begin{equation} \lim_{\nu \to \infty}
\sum_{i=0}^\nu C_i R_i^D = C_0 R_0^D \lim_{\nu \to \infty} \sum_{i=0}^\nu \left( {\displaystyle
1+ {\displaystyle n_{D,0} \over n_{0,D}}\over x^D}\right)^i = A \;\;\; .  \label{Volume}
\end{equation} This condition gives a lower limit for $ x $:  \begin{equation} x> \left(1+ {
\displaystyle n_{D,0}\over n_{0,D} } \right)^{1/D} \;\;\; .  \label{low_x} \end{equation}

An upper limit for $x$ can be calculated solving the algebraic equation obtained substituting $
\epsilon_{\nu}$ with $\epsilon_\nu=x^\nu$, into the generalization to any $D$ of
eq.(\ref{bends}) \cite{Goss37}.  (For example one obtain $x \simeq 1.88$ for $D=3$, $x \simeq
1.55$ for $D=4$, $x \simeq 1.39$ for $D=5$, $x \simeq 1.30$ for $D=6$ and $x=1$ for $D
\rightarrow \infty$).

The size distribution can be obtained by associating eq.(\ref{recur}) with the condition
$R_{\nu+1}/R_\nu=x$.  One has \begin{equation} N(R_\nu)=N_0 \left( {\displaystyle R_0 \over
R_\nu} \right)^\alpha \;\;\; , \label{distr} \end{equation} where the coefficient is
\begin{equation} \alpha={\displaystyle \ln(1+{n_{D,0} \over n_{0,D}}) \over \ln x} \;\;\;\;.
\label{expon} \end{equation}

By substituting the lower limit of $x$ (maximum packing, eq.(\ref{low_x}) ) one obtains the
upper bound for the exponent $\alpha_{Max}=D$.  The lower bound for $\alpha$ can be calculated
substituting into eq.(\ref{expon}) the maximum value of $x$ (that corresponds to the Apollonian
case) and the minimum value of the coefficient ${n_{D,0}/ n_{0,D}}$.  This last parameter
depends on the packing strategy.  For $D=3$, the Apollonian packing has:  $n_{0,D}=4$,
$n_{D,0}>12$ and $x \simeq 1.88$.  In this case, one gets the limit of $\alpha \simeq 2.2$.

The two models, discussed in app.\ref{A.2} can be easily extended to the $3D$ case, one obtains
$\alpha=1.26$ and $\alpha=2$ for the hexagonal and triangular models respectively.

\section{\label{S.4} Computer simulations}

The covering procedure starts positioning the circles with maximum radius $R_0=R_{Max}$.  The
position of the centre of each circle is chosen at random.  A new circle is added to the system
only if it is non overlapping with any other pre-existing circle, otherwise another random
position is chosen.  This first step ends when all the interstitial spaces between circles have
sizes smaller than the circle diameter and thus new circles with radius $R_{Max}$ cannot be
positioned any more.  At the second step, the radius of the new circles is reduced$R_1=R_0/x$,
the centre are chosen at random and a new circle is placed only if it is non overlapping with
any other pre-existing circle.  The procedure continues gradually reducing the sizes of the
circles to place once the interstitial spaces become too small to contain new circles.  No
shifting or rearrangement are performed once a circle have been positioned.  The procedure ends,
after a finite number of steps, at the lowest radius $R_{min}$.

\begin{figure}
\vspace{-6.cm}
\hspace{1.cm}
\epsfxsize=16.cm
\epsffile{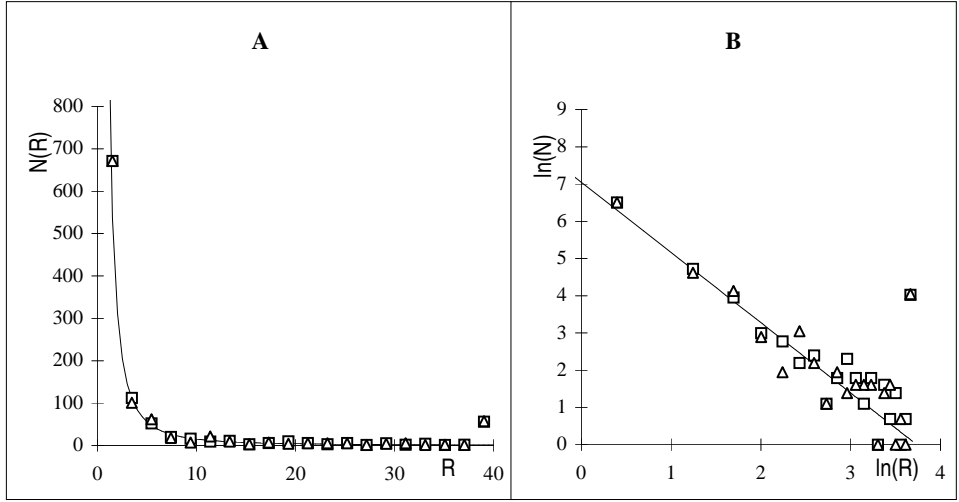}
\label{f.3}
\vspace{-10.cm}
\caption{Circle-size distributions of two computer generated circle coverings. (a) Normal scale.
(b) Double logarithmic scale. The full line is the average best fit. 
}
\end{figure}

Figures
(\ref{f.3}a,b) show the circle sizes distribution in normal and double logarithmic scale, for
two simulations.  The data have been obtained covering a rectangle of $600 \times 450$ pixels
with periodic boundary conditions.  The maximum radius was $R_{Max}=45$ pixels and the minimum
was $R_{min}=1.5$ pixels.  In accordance with the covering procedure, the circle size have been
decreased between $R_{Max}$ to $R_{min}$ in $50$ finite steps.  In the simulations reported in
fig.(\ref{f.3}), the total number of circles was respectively equal to 995 (squares) and 987
(triangles).

The circle size distribution is characterized by one peak at the maximum radius and by a fast
increment towards the minimum radius.  The linear trend of the distribution, in the region of
small radii ($R<R_{Max}/5$), shown in the double logarithmic scale (fig.(\ref{f.3}b) ), suggests
a power law $N(R)\propto R^{-\alpha}$.  The best-fit estimation for the exponent give $\alpha=
1.84 \pm 0.16$ (squares) and $\alpha = 1.92 \pm 0.19$ (triangles), with confidence factors
respectively equal to $93\%$ and $92\%$.

A power law for the sizes distribution was analytically predicted in the previous paragraphs and
in app.\ref{A.1}.  The best--fit values of the exponents $\alpha$ are in good agreement with the
theoretical predictions for the case of disordered maximum circle--packing.

\section{\label{S.5} Structures formed by condensed tin droplets} 

\begin{figure}
\vspace{-1.cm}
\hspace{-4.cm}
\epsfxsize=26.cm
\epsffile{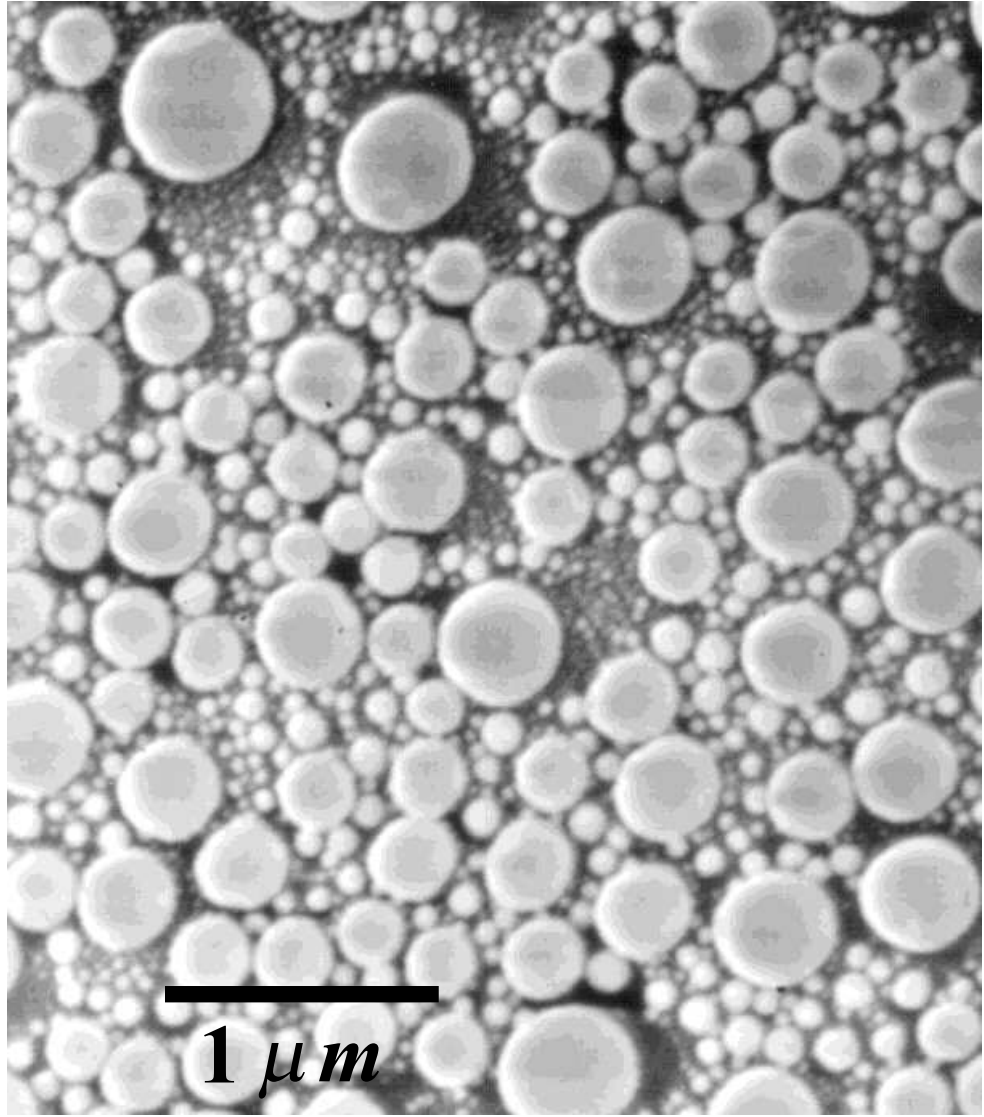}
\label{f.4}
\vspace{-18.cm}
\caption{SEM micrograph of a ``breath figure'' generated by tin drops deposited 
on a hot alumina substrate.}
\end{figure}

\begin{figure}
\vspace{-2.cm}
\hspace{-2.cm}
\epsfxsize=22.cm
\epsffile{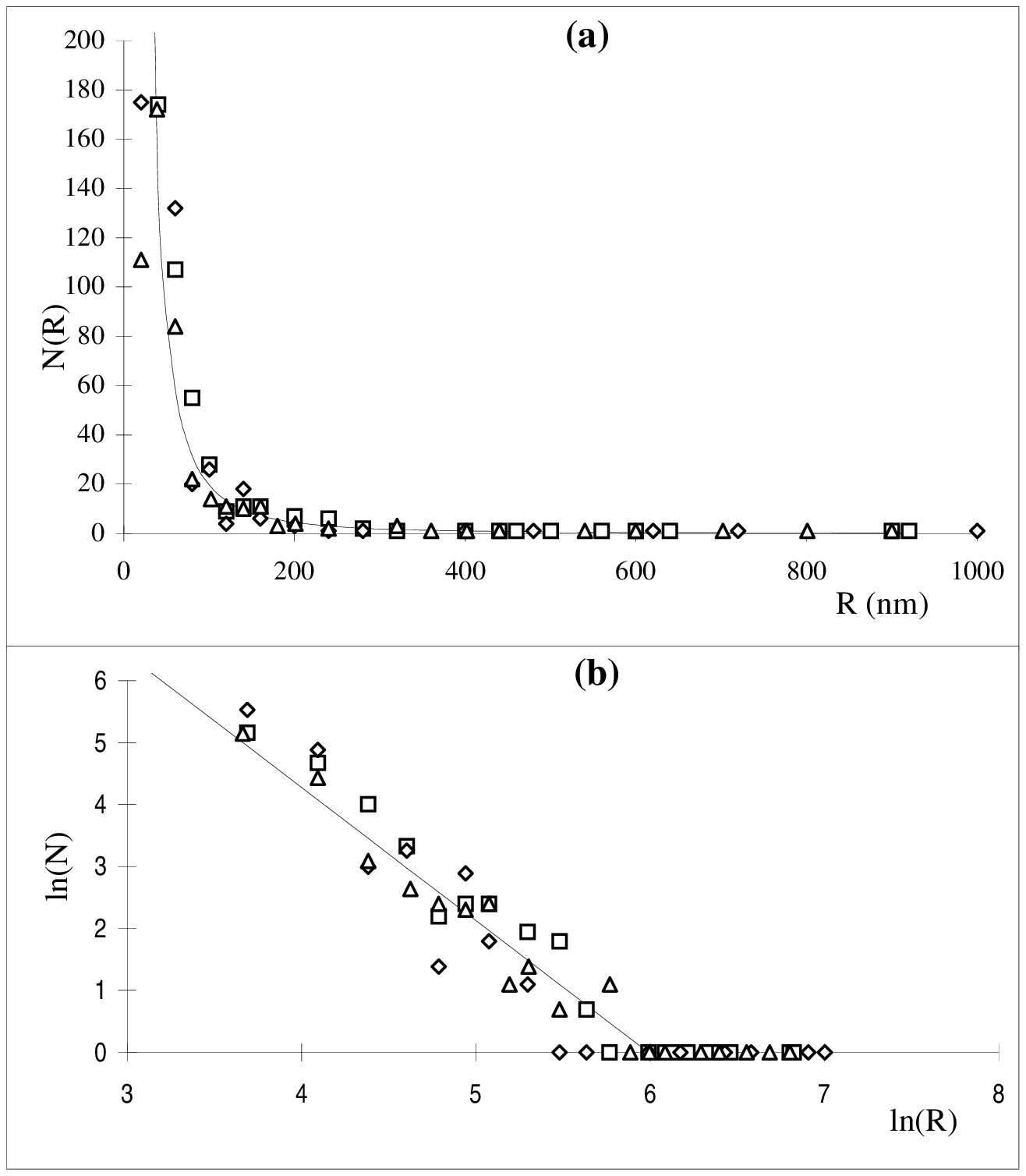}
\label{f.5}
\vspace{-10.cm}
\caption{Size distribution of the tin drop diameters for three different samples.
(a) normal scale; (b) double logarithmic scale.
In ordinates, N(R) is the number of drops per $\mu m^2$. 
The full line is the average best fit.}
\end{figure}

Figure (\ref{f.4}) is a SEM
micrograph of Sn drops condensed on a hot, flat alumina substrate.  The system have been
prepared evaporating the tin in high vacuum on a substrate heated at a temperature higher than
the tin melting point \cite{Sberv,Aste/Bot94}.  After cooling at room temperature, one has a
stable system of packed tin drops.  The size-distribution of such a system was studied using SEM
micrographs of magnifications 10 and 20 $K\propto$.  The micrographs were digitalized with a
scanner.  On each image an internal rectangular perimeter was defined and the drops inside and
on the bound perimeter were encircled with circles of the same size of the surface occupied by
the drop.  The size distribution of these circles was studied.  The boundary effects were taken
into account counting 1/2 the circles crossing an edge of the perimeter and counting 1/4 the
circles containing a vertex of the bound-perimeter.  In fig.(\ref{f.5}) the sizes-distributions
for three different samples are reported in normal and double logarithmic scales.  The three
samples were deposited respectively with 2.5 $g/m^2$ (triangles), 2.8 $g/m^2$ (squares) and 3.2
$g/m^2$ (diamonds).  The biggest drops have sizes of about $1000 nm$, whereas the smallest
(measurable) drop-sizes are $50 nm$.  The number of drops encircled per each micrograph was
about 2000-2500.  The distribution is characterized by rather uniform drop sizes in the region
of large radii and by a fast increment, as a power law, towards the smallest radii.

Typically, one drop nucleates on a preferential centre and grows supported by condensation.
This independent growth stops when the drop touch another surrounding drop.  At this point, the
two drops melt together and the formed new drop --eventually-- melts with other surrounding
drops in a chain of coalescence events.  The coalescence leads to a new drop with volume equal
to the sum of the volumes involved in the coalescence phenomena.  This new drop occupies a
surface lower than the sum of the surfaces occupied by the original drops.  It follows that the
coalescence mechanism liberates space on the substrate.  On this surface the nucleation, growth
and coalescence phenomena can start again.  The final structure is therefore a consequence of a
very complicated process which involve many body phenomena \cite{Bey86}.  Nevertheless the
formation of the final structure can be investigated adopting a simple point of view.

The system of quenched drops showed in fig.(\ref{f.4}) is apparently similar to the systems
studied, analytically and numerically, in the previous paragraphs.  This similarity is not only
apparent:  the nucleation, growth and coalescence mechanism leads to systems where the drops
occupies the maximum possible surface and have the biggest sizes, compatibly with the
coalescence that prevents the drops overlapping.  The system evolves through self--similar
configurations where the size distribution is, at any time, the maximum packing distribution.

The similarity between this two systems is confirmed by the drop size distribution:  in the
region of little radii the distribution follows a power law $N(R) \propto R^{-\alpha}$ with
exponent $\alpha \simeq 2$.  The best--fit give $\alpha = 1.96 \pm 0.13$ (triangles in
fig.(\ref{f.5}) ), $\alpha = 2.05 \pm 0.11$ (squares) and $\alpha = 2.1 \pm 0.23$ (diamonds),
with confidence factors respectively equal to $94\%$, $95\%$ and $87\%$.  These values are in
accordance with the analytical predictions and strongly suggests that the system morphogenesis
is ruled by the mechanism of the maximum circle-packing.

In the region of big radii ($R\simeq R_{Max}>500\; nm$) the drops have a rather uniform sizes.
This region of uniform sizes is the memory of the first stage of the deposition where the drops
grow independently.  Indeed, in this stage the relative volumes of the drops are proportional to
the sizes of the Voronoi cell constructed around the drop centers \cite{Mul95}, and a rather
uniform distribution is expected.  In fig.(\ref{f.5}) this uniformity at $R \simeq R_{Max}$ is
not particularly manifest since the number of drops in this region is little in comparison with
the number of small drops ( for example in a sample deposited with about 3 $g/m^2$, one has
about 10 drops per $\mu m^2$ of about $0.1 \mu m$ and 1 drop per $\mu m^2$ of $1 \mu m$).
Despite the fact that their number is little, the big drops cover a large amount of surface and
contain the main part of the volume deposited (using the same example reported above, it is
straightforward to see that the big drops cover an area 10 times larger and occupy a volume 100
times larger than the small drops).  Note that, a peak at $R_{Max}$ is present also in the
computer simulations.  A deviation from the maximum packing law in the region close to $R_{Max}$
is not in contradiction with the theoretical predictions.  The analytical result $N(R) \propto
1/R^\alpha$, concerns only the asymptotic limit of the distribution ($\nu \rightarrow \infty$
i.e.  $R/R_{Max} \rightarrow 0$) where the system is supposed to be scale--invariant.

Studies reported in literature \cite{Bey86,Bri90,Fri90,Kno88,Fam88,Derr91,Bey90I,Fri88}, show
that the system of condensed drops evolve in time increasing the drop sizes, decreasing the
drops number and maintaining constant the coverage.  In the systems of tin drops here studied,
the coverage have been experimentally found in the range 60\% to 65\% \cite{PozTes},
\cite{Aste/Bot94}.  A small value of the coverage (typically in breath figures is equal about
55\% \cite{Derr91}) is a consequence of coalescence that continuously liberates space.  Consider
a configuration of three external drops of radius $R_0$ and one interstitial drop of radius
$R_1$.  The collapse of these four drops into one trough coalescence liberates a certain amount
of surface.  The fraction of area liberated (Area occupied by the drops after coalescence/ Area
occupied before coalescence) is slightly dependent on the packing strategy.  It is equal to
$32\%$ if the four drops are all ``kissing'' each other and equal to $35\%$ in the case of
disordered maximum packing ($R_0/R_1=x \simeq \sqrt 3$).  The presence of other interstitial
drops that participate to the coalescence phenomena reduces the previous values within $1\%$.
Starting with an initial --hypothetical-- coverage of $100\%$, the coalescence reduces the
coverage to $68\%$ and $65\%$ in the two cases discussed above.  These values are close to the
coverage experimentally observed.  Note that these values are scale invariant and depends only
on the local configurations of drops (i.e.  on the packing strategy).  The scale invariance
implies that the fraction of area liberated by coalescence is independent of the drops radii
(i.e.  independent of the amount of tin deposited) thus constant during the deposition.

\section{Conclusions}

Circle--covering is a good model--system for many two dimensional natural cellular system where
the plane is filled in the most efficient way compatibly with the cell shapes and sizes.  It has
been found (\S \ref{S.2}, \S \ref{S.3} and app.\ref{A.1}) that polydisperse circles packed in a
dense way, have a size distribution that follows the general law $N(R) \propto R^{-\alpha}$.
This distribution is a consequence of the scale--invariance in the packing strategy.  The range
of variability of the exponent have been calculated.  We have obtained a maximum value of
$\alpha_{Max}=2$ for the general disordered case, where the circles are arranged at random
following the only constraint of non overlapping.  The minimum value has been estimated
$\alpha_{min} \simeq 1$ and corresponds to an approximate solution for the Apollonian packing
(\S \ref{S.2.1} and \S \ref{S.3}), and to an exact solution for the ``hexagonal'' filling model
(app.\ref{A.2}).

These results, obtained for two dimensional circle packing, have been extended to packing of
spheres in spaces of arbitrary dimension (\S \ref{S.3.1}).  We found that, in $D$--dimensional
spaces, the size distribution of densely packed $D$--spheres follows --as in 2D-- a power law
$N(R) \propto R^{-\alpha}$.  In this generalized case, the maximum value of the exponent is
$\alpha_{Max}=D$ and is associated with disordered packings.  The minimum value is associated
with the packings of tangents $D$--spheres.  For $D=3$, this value have been estimated equal to
$\alpha=2.2$ for the Apollonian case, whereas for the hexagonal filling model we found
$\alpha=1.26$.

Computer simulations of two--dimensional circle--packing confirm the analytical predictions:  in
the disordered case the size--distribution follows a power law with exponent $\alpha \simeq 2$
(see \S \ref{S.4}).

The formation of breath figures have been interpreted in terms of circles packing regulated by
coalescence (\S \ref{S.5}).  The mechanism of formation of this figures is very complicated
since it is a dynamic system where the evolution involves many--body phenomena.  On the other
hand, the structures formed by the drops instant by instant have strong similarities with the
structures generated by packing circles.  In particular, the size distribution in the region of
little radii, follows the same power law as the disordered, dense circle-packing with exponent
$\alpha \simeq 2$.  The coverage, evaluated through the fraction of area liberated by
coalescence in a system of drops that follows the maximum packing distribution, is in agreement
with the experimental observation.

\vskip1.5cm {\large \bf Acknowledgments} \vskip.7cm \noindent The author wish to thank Rodolfo
Botter for the help given for the samples preparation, the microstructural characterization and
for the useful discussions.  The author thanks F.  Saya and P.  Pozzolini for the measurements
of the drop--sizes distributions.  Part of the ideas reported in this paper have been originally
developed in the author's PhD thesis which have been written under the bright supervision of
Prof.  D.  Beruto.  Finally, the author thanks Nicolas Rivier for the useful discussions.

\vskip3.cm

\appendix \section{\label{A.1} Size-distribution and covering strategies}

In the second and third paragraphs we found that the size distribution of densely packed circles
follows a power law (eq.(\ref{distrib}) ) where the exponent $\alpha$ is related to the
topological properties of the Dodds network and to the parameter $x$ (eq.(\ref{exponent}) ).
This is a general result where the only a-priori hypothesis is the existence of a Dodds network
with convex cells (this condition can be considered as the definition of dense packing).  In
this appendix we generalize these results using an approach which doesn't needs the definition
of the Dodds network.

Consider a covering procedure which fills the available space with non overlapping circles
starting from the circles of bigger sizes $R_{Max}$ and then gradually reducing the sizes
$R_\nu<R_{Max}$ in order to fill the interstitial spaces.  The number of circles introduced at
any covering step and the relative sizes depend on the covering procedure.  In the second
paragraph we pointed out that the scaling condition implies that the number of circles inserted
at any covering step grows as a geometrical progression $N_\nu \propto a^\nu$ and the size of
the circles decreases as a geometrical progression $R_\nu \propto R_0/ x^\nu$.  As a consequence
the size distribution results a power law $N(R_\nu) \propto R_\nu^{-\alpha}$ with exponent
\begin{equation} \alpha={\displaystyle \ln a \over \ln x}\;\;\;\;\; .  \label{alpha}
\end{equation} In the case of dense packing (i.e.  when it is possible to define a Dodds network
with convex cells) we found $a=\langle z \rangle/2$ (eq.(\ref{expon}) ) and the bound $x >
\sqrt{ \langle z \rangle/2}$ (eq.(\ref{xx}) ).

More generally one can construct the Delaunay triangulation with the centre of the circles as
vertices.  This triangulation is always well defined and (for an infinite system) the average
connectivity is $\langle z \rangle = 6$.  Different packing strategies differentiates in the
number, the sizes and the positions of the new vertices (i.e.  the new circles) to insert.  From
a topological point of view, there are only two position where the centre of the new circles can
be placed:  inside a triangle or on an edge (the centre on a vertex is forbidden by the
condition of non overlapping).  Suppose that the packing procedure places a new circle with
probability $p$ inside a given triangle and with probability $q$ on a given edge of the Delaunay
triangulation.  If $V_\nu$ is the number of vertices (i.e.  of circles) at the covering stage
$\nu$, it is easy to prove (following the same arguments of \S\ref{S.2}) that at the next stage
this number is \begin{equation} V_{\nu+1}= (1+2p+3q)V_\nu\;\;\;.  \label{V} \end{equation} From
relation (\ref{V}) follows immediately that the number of circles grows as a geometrical
progression ($N_\nu=V_\nu \propto a^\nu$) with coefficient $a=(1+2p+3q)$.

The argument used in \S\ref{S.2} to find the lower bound on the parameter $x$ (eq.(\ref{xx}) )
can be directly extended to the present case.  One gets $x>\sqrt a$, which, substituted in
(\ref{alpha}), gives the upper limit $\alpha_{Max}= 2$.

The Apollonian covering is a particular example of the covering procedure here discussed.  In
this case one has $p=1$, $q=0$ and $x \simeq 2.9$, which gives $\alpha \simeq 1$.

\section{\label{A.2} Hexagonal and triangular filling models}

Following the work of Bidaux {\it et al.}  \cite{Bid73}, let us describe two simple
circle-covering problems where the exponent $\alpha$ can be calculated exactly.  Originally
these models have been proposed as simplified geometrical problems to estimates two bounds for
the fractal dimension in the Apollonian packing.  In the present work these two models are
presented as good examples of circle-covering which do not necessarily corresponds to
approximations of the Apollonian case.

Let us start with one triangle.  In the first model ({\it triangular}), one insert a new
triangle with vertices in the mean point of the edges of the original triangle.  In this way the
original triangle is divided in 4 identical triangles which are similar to the original.  Then,
one insert a circle inside the central triangle and iterate the procedure on the three external
triangles.  At the beginning one starts with 1 circle, at the first step one insert 3 new
circles, at the $\nu^{th}$ step the number of new circles inserted is $3^\nu$.  The parameter
$a$ (see.  app.\ref{A.1}) is then equal to 3.  The ratio $x$ can be easily derived by observing
that the triangles are all similar and that at each step of the sequence the sizes of the edges
are reduced by a factor 2.  Consequently one has $x=2$.  Substituting the parameters $a$ and $x$
into eq.(\ref{alpha}) one gets the exponent $\alpha = \ln 3 / \ln 2 \simeq1.58$.

In the second model ({\it hexagonal}) one inscribe first an hexagon with vertices which are
dividing in three equal part the edges of the original triangle.  Then a circle is inscribed
inside the hexagon and the procedure is iterated on the three remaining external triangles.  As
before one obtains $a=3$, whereas the scale factor is $x=3$.  By substituting into
eq.(\ref{alpha}) we get the exponent $\alpha=1$.

Note that, if one use equilateral triangles the coverage $\theta$ (Area covered by the circles/
Area of the original triangle) can be calculated exactly.  One has $\theta = \pi/(3 \sqrt 3 )
\simeq 0.6$ for the triangular model and $\theta = \pi/(2 \sqrt 3 ) \simeq 0.9$ for the
hexagonal.  These value can be increased by inserting new triangles in the free interstices
between the circles and iterating the procedure.

\end{document}